\documentclass[letterpaper,journal]{IEEEtran}
\usepackage{amsmath,amsfonts}
\usepackage{algorithmic}
\usepackage{algorithm}
\usepackage{array}
\usepackage[caption=false,font=footnotesize,labelfont=sf,textfont=sf,skip=2pt]{subfig}
\usepackage{textcomp}
\usepackage{stfloats}
\usepackage{url}
\usepackage{verbatim}
\usepackage{graphicx}
\usepackage{cite}
\usepackage{siunitx}
\usepackage{multirow}
\usepackage{booktabs}
\usepackage{makecell}
\usepackage{enumitem}
\usepackage{colortbl}
\usepackage{xcolor}
\usepackage{soul}
\sethlcolor{yellow}

\usepackage{float}
\begin{document}

\title{Accurate Frequency Response Modeling in Integrated T\&D Co-Simulation via EWMA-RTTA-Based Quadratic Extrapolation}

\author{Jong Ha Woo,
        Qi Xiao,
        Yu Ma,
        Zishuo Yang,
        Victor Daldegan Paduani,
        and Ning Lu
\thanks{J. H. Woo, Q. Xiao, Y. Ma, Z. Yang, and N. Lu are with the Department of Electrical and Computer Engineering, North Carolina State University, Raleigh, NC 27606, USA (e-mail: jwoo@ncsu.edu; qxiao@ncsu.edu; yma5@ncsu.edu; zyang22@ncsu.edu; nlu2@ncsu.edu).}%
\thanks{V. D. Paduani is with the New York Power Authority, White Plains, NY 10603, USA (e-mail: victor.paduani@nypa.gov).}%
\thanks{This work has been submitted to the IEEE Transactions on Power Systems for possible publication. Copyright may be transferred without notice, after which this version may no longer be accessible.}}

\markboth{Preprint -- Submitted to IEEE Transactions on Power Systems, \today}%
{Woo \MakeLowercase{\textit{et al.}}: Accurate Frequency Response Modeling in Integrated T\&D Co-Simulation via EWMA-RTTA-Based Quadratic Extrapolation}


\maketitle

\begin{abstract}
The large-scale integration of inverter-based resources (IBRs), particularly distributed photovoltaics (DPVs), into distribution networks increases the need for integrated transmission and distribution (T\&D) co-simulation. A key challenge in such co-simulation lies in accurately modeling system frequency across two asynchronous simulation environments. For example, the transmission system, simulated in the phasor domain, can operate with a simulation timestep of 10~ms, while the distribution system, simulated in the electromagnetic transient domain (EMT) to include IBR models, uses a much finer timestep of \SI{100}{\micro\second}.
To ensure accurate PLL-based frequency estimation in distribution systems, it is essential to predict voltage magnitude and phase angle variations within the 10~ms transmission intervals, rather than using constant values that cause inaccurate frequency calculations. This issue becomes particularly critical when modeling primary and secondary frequency response services provided by IBRs.
To address this challenge, we propose an automated \textit{Exponentially Weighted Moving Average Real-Time Threshold Adaptation} (EWMA-RTTA) method, which utilizes \textit{Quadratic Extrapolation} to predict voltage magnitude and phase angle trends more precisely. The proposed method is validated using two Opal-RT simulators: one simulating an IEEE 118-bus transmission system and the other simulating an IEEE 123-bus distribution network. Simulation results demonstrate that our approach improves the normalized mean absolute error (nMAE) by a factor of 25.7 compared to methods that do not account for time mismatches, offering a scalable and accurate solution for modeling IBR-based frequency response in modern power systems.
\end{abstract}


\section*{Nomenclature}
\begin{description}[labelwidth=1.5cm, leftmargin=1.7cm, font=\normalfont]
\item[$T_T$] Transmission system timestep.
\item[$T_D$] Distribution system timestep.
\item[$V_t$] Voltage magnitude at time $t$.
\item[$\theta_t$] Voltage phase angle at time $t$.
\item[$f_t$] Frequency at time $t$.
\item[$f_0$] Nominal system frequency (60 Hz).
\item[$\alpha_t$] EWMA smoothing factor at time $t$.
\item[$\mu_t$] EWMA mean at time $t$.
\item[$\sigma_t$] EWMA standard deviation at time $t$.
\item[$TH_t$] Time-varying threshold at time $t$.
\item[$\Delta y_t$] Data change at time $t$.
\item[$\mathbf{w}$] Weight vector for extrapolation.
\item[$\mathbf{y}$] Vector of past data points.
\end{description}

\section*{Acronyms}
\begin{description}[labelwidth=2.2cm, leftmargin=2.4cm, font=\normalfont]
\item[AGC] Automatic Generation Control.
\item[DPV] Distributed Photovoltaic.
\item[EMT] Electromagnetic Transient.
\item[EWMA] Exponentially Weighted Moving Average.
\item[EWMA-RTTA] Real-Time Threshold Adaptation using EWMA.
\item[IBR] Inverter-Based Resource.
\item[LPF] Low-Pass Filter.
\item[MAPE] Mean Absolute Percentage Error.
\item[nMAE] normalized Mean Absolute Error.
\item[PCC] Point of Common Coupling.
\item[PFR] Primary Frequency Response.
\item[PLL] Phase-Locked Loop.
\item[RTS] Real-Time Simulator.
\item[SFR] Secondary Frequency Response.
\item[T\&D] Transmission and Distribution.
\end{description}

\section{Introduction}
\IEEEPARstart{T}{}he rapid deployment of inverter-based resources (IBRs), particularly distributed photovoltaics (DPVs), has transformed modern power systems, introducing challenges in maintaining power balance and frequency stability across transmission and distribution (T\&D) networks \cite{ref1}. Driven by declining costs and supportive policies such as FERC Order 2222, DPVs have become integral to sustainable energy systems \cite{ref2,ref2_2}. However, their reliance on power electronic inverters—lacking the inherent inertia of synchronous generators—results in faster frequency deviations during disturbances \cite{ref3}. With FERC Order 842 (2018) mandating that DPVs provide Primary Frequency Response (PFR), accurate modeling becomes critical for frequency stability \cite{ref4}.

T\&D simulation methods are generally classified into \textit{integrated single-system} simulations and \textit{distributed} simulations. The latter approach, commonly known as co-simulation, involves asynchronous information exchange between separate simulators. Integrated single-system simulations model both transmission and distribution within a unified framework, providing comprehensive analysis but incurring high computational costs that limit scalability for large-scale systems \cite{ref4}. These simulations require uniform time steps (e.g., 1 ms) to capture fast dynamics, which are computationally prohibitive for complex transmission networks \cite{ref5}.

In contrast, distributed co-simulation employs separate simulators. Early approaches use static modeling for both transmission and distribution, such as Matpower and OpenDSS for optimal power flow analysis, but cannot capture system dynamics \cite{ref5,ref6,ref7,ref8}. Other frameworks combine dynamic transmission modeling using tools like RTAG, PowerWorld, PSSE, ANDES, or PSAT with static distribution modeling using OpenDSS \cite{ref9,ref9_2,ref10,ref11,ref12,ref13}. However, reliance on static distribution models lacks the differential equations required for dynamic response modeling, leading to significant errors in DPV frequency response evaluation due to \textit{discrepancies between the system frequency and the frequency at the point of common coupling (PCC)} \cite{ref12}.

To address the limitations of static distribution models, some frameworks adopt dynamic modeling for both transmission and distribution networks, applying \textit{uniform time steps} (e.g., $T_T = T_D \leq 1 \, \text{ms}$) using simulators like GridPack, InterPSS and Real-Time Simulators (RTS) like OPAL-RT, or RTDS \cite{ref14,ref15,ref16,ref17}. However, this approach becomes computationally prohibitive for large-scale transmission networks with over 100 generators, where slow electromechanical dynamics do not require fine temporal resolution.

\IEEEpubidadjcol

Alternative frameworks improve computational efficiency by adopting \textit{asynchronous time-stepping}, where transmission networks use larger time steps $T_T$ (e.g., 10 ms) while distribution networks use smaller time steps $T_D$ (e.g., \SI{100}{\micro\second}) to capture fast inverter-based dynamics \cite{ref18,ref19}. However, existing studies have largely ignored the effects of timestep mismatches ($T_T \neq T_D$). The rate transition methods commonly used lack extrapolation capability and simply repeat voltage magnitude and phase angle data across multiple distribution timesteps. This repetition can cause Phase-Locked Loops (PLLs) to misestimate frequency, leading to significant errors in frequency response. In utility power system simulations, such inaccuracies could result in misguided evaluations of contingency scenarios, leading to inadequate planning and heightened risks of real-world grid failures \cite{ref26}.

\begin{table*}[!t]
\centering
\fontsize{8}{8}\selectfont
\renewcommand{\arraystretch}{1.5} 
\caption{Comparison of T\&D Co-Simulation Frameworks}
\begin{tabular}{>{\centering\arraybackslash}m{2.5cm} @{\hspace{0.5em}} >{\raggedright\arraybackslash}m{2.5cm} @{\hspace{0.5em}} >{\centering\arraybackslash}m{1.5cm} @{\hspace{0.5em}} >{\raggedright\arraybackslash}m{2.5cm} @{\hspace{0.5em}} >{\centering\arraybackslash}m{1.5cm} @{\hspace{0.5em}} >{\raggedright\arraybackslash}m{6.5cm}}
\toprule[1.5pt]
Reference & T. Simulator & $T_T$ & D. Simulator & $T_D$ & Characteristics \\
\midrule[1.5pt]
{[5]} & ISM & 1 ms & ISM & 1 ms & Integrated T\&D within a single simulator is not applicable in large complex T\&D networks \\ \hline
{[6]--[9]} & Matpower & -- & OpenDSS & -- & T\&D both static, cannot observe any system dynamics \\ \hline
{[10]--[15]} & RTAG  \newline ANDES \newline PSAT & 10 ms & OpenDSS & -- & T: Dynamic, D: Quasi-static, exchange interval $\geq 2$ s, misses fast transients and inaccurately assesses IBR responses \\ \hline
{[16]--[19]} & OPAL-RT \newline RTDS & $T_T \leq$ \newline 1 ms & OPAL-RT \newline CDP & $T_D \leq$ \newline 1 ms & Uniform time steps ($T_T = T_D$), but small time steps are challenging for large-scale systems \\ \hline
{[20]--[21]} & GridPack \newline TSAT & $T_T \leq$ \newline 10 ms & GRIDLAB \newline RTDS & $T_D \leq$ \newline 10 ms & Different time steps ($T_T \neq T_D$) cause errors in PLL frequency calculation at PCC \\ \hline
\noalign{\vskip\arrayrulewidth}\hline\noalign{\vskip\arrayrulewidth}
Our approach & RTS (OPAL-RT) & 10 ms & RTS (OPAL-RT) & \SI{100}{\micro\second} & Even with different time steps (\( T_T \neq T_D \)), extrapolation enables accurate calculation of PCC frequency and evaluation of Frequency Response \\
\bottomrule[1.5pt]
\end{tabular}
\label{tab:comparison}
\vspace{-1em}
\end{table*}

To address this challenge, this paper proposes a novel framework that integrates Real-Time Threshold Adaptation using an Exponentially Weighted Moving Average (EWMA-RTTA) with quadratic extrapolation. The proposed approach resolves data resolution mismatches and enables accurate evaluation of DPV frequency responses across systems with different time steps.

As summarized in Table~\ref{tab:comparison}, the main contributions of this work are as follows:
\begin{enumerate}
    \item We propose a real-time \textit{quadratic extrapolation} method to predict transmission-side nodal voltage magnitude and phase angle variations between transmission updates. This enables the distribution system to generate smooth, accurate three-phase voltage waveforms for PLL-based frequency estimation at \SI{100}{\micro\second} resolution using transmission data updated every 10~ms. The method achieves low mean absolute percentage errors of 0.0104\% for voltage magnitude, 0.307\% for phase angle, and 0.000563\% for PLL frequency.
    \item We develop an \textit{exponentially weighted moving average real-time threshold adaptation} (EWMA-RTTA) mechanism to detect abrupt data changes (e.g., generator trips) and dynamically adjust thresholds. The proposed approach is robust to large timestep mismatches (e.g., 100$\times$) and minimizes error propagation, enabling accurate frequency response evaluation of inverter-based resources such as DPVs.
\end{enumerate}

Experimental validation was performed using the IEEE 118-bus transmission system and the IEEE 123-bus distribution system, implemented on two real-time simulators and coupled via Python-based TCP/IP and Modbus communication~\cite{ref20}. 

Compared to the baseline, the proposed framework reduces the normalized mean absolute error (nMAE) in PV output modeling by nearly 30$\times$ (from 0.278\% to 0.0108\%), enabling accurate and scalable DPV frequency response modeling in high-IBR power systems. To the authors' knowledge, this is the first study to jointly address timestep mismatch resolution in T\&D co-simulation and accurate DPV frequency response modeling.

Notably, secondary frequency response is not always applicable to inverter-based resources (IBRs), whereas local control–based primary frequency response (PFR) is both practical and essential. Accordingly, our methodology prioritizes accurate exchange of voltage magnitude and phase angle data to support PFR as the core application, ensuring robust performance in high-IBR power systems.

The remaining sections are organized as follows: Section II presents the methodology, Section III details the simulation results, and Section IV concludes with future research directions.

\begin{figure*}[!t]
\centering
\includegraphics[width=1.8\columnwidth]{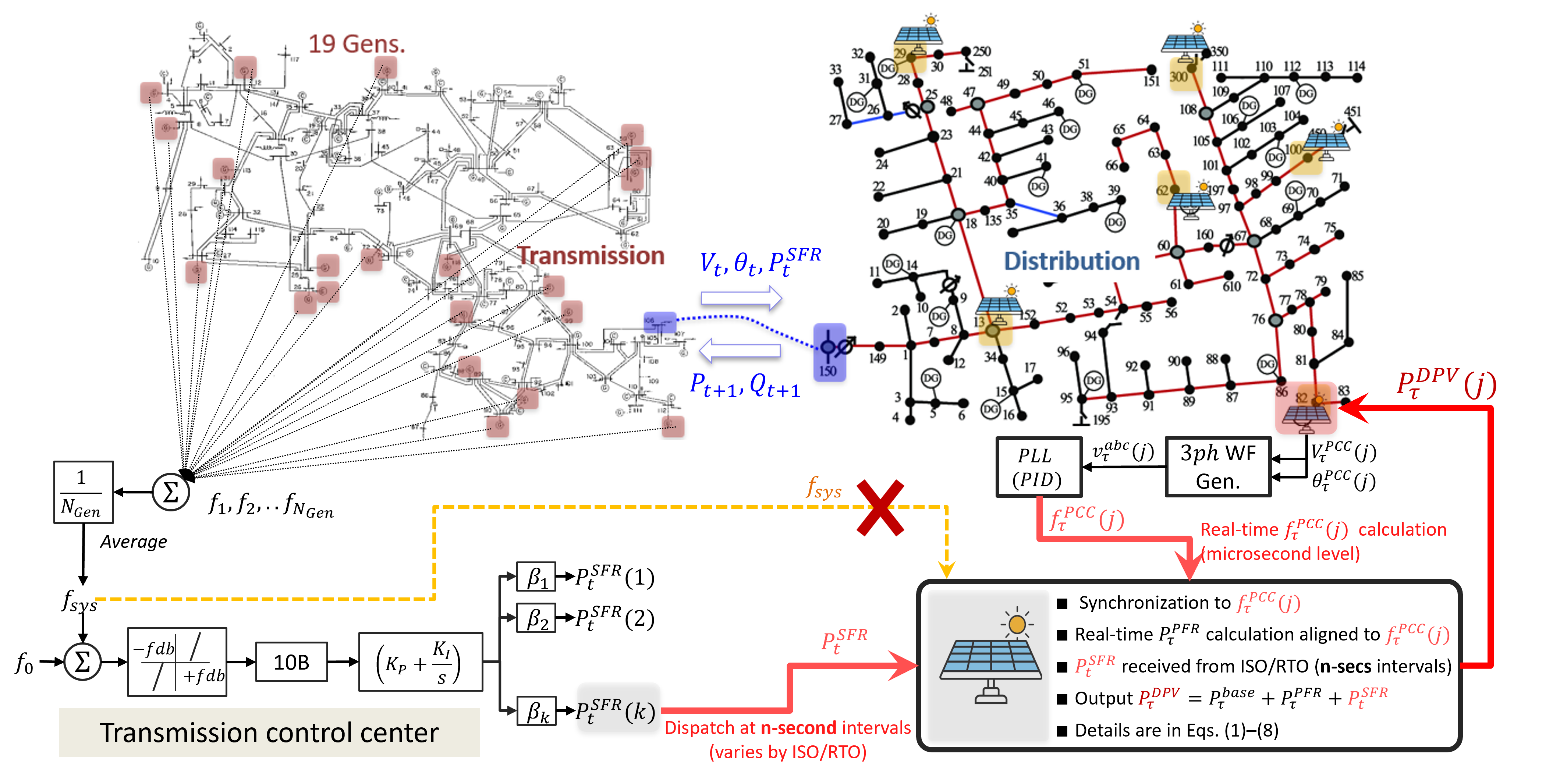}
\caption{Workflow of PV frequency response in T\&D co-simulation}
\label{fig:pfr_workflow}
\vspace{-1em}
\end{figure*}

\section{Methodology}
As illustrated in Figure~\ref{fig:pfr_workflow}, modeling the frequency response provided by IBRs requires an integrated transmission and distribution (T\&D) co-simulation framework. The transmission network uses the IEEE 118-bus system in phasor-domain, while the distribution network employs the IEEE 123-bus system in phasor-domain with EMT-domain IBR models for DPVs.

The transmission system, solved with timestep \( T_T \) (e.g., 10~ms), yields bus voltage magnitude \( V_t \) and phase angle \( \theta_t \). At transmission bus \( i \) connected to the distribution network, local voltage magnitude \( V_t(i) \), angle \( \theta_t(i) \), and secondary frequency response power \( P_{\mathrm{SFR}(i)} \) are transferred to the distribution feeder head. The IEEE 123-bus distribution model runs at finer timestep \( T_D \) (e.g., 100~$\mu$s), using the received voltage magnitude and angle from the transmission system to establish boundary conditions at the feeder head, while simulating DPV behavior at node \( j \) (PCC).

The DPV system generates output power (\( P^{ref}_{\mathrm{t+1}} \)) as the sum of \( P^{base}_{\mathrm{t+1}} \), \( P^{PFR}_{\mathrm{t+1}} \), and \( P^{SFR}_{\mathrm{t+1}} \). The total active and reactive power consumption (\( P_{\mathrm{t+1}}, Q_{\mathrm{t+1}} \)) of the entire distribution network, measured at the distribution feeder head, are fed back to the transmission network at bus \( i \) to update the load bus (PQ bus) conditions for coordinated T\&D operation.

\vspace{-1em}
\subsection{Effects of Time Step Mismatches}
Generator buses allow direct frequency measurement through rotor angle dynamics. In contrast, for distribution node \( j \) serving as the PCC, local frequency \( f_t(j) \) is estimated using a Phase-Locked Loop (PLL). The transmission system outputs voltage magnitude \( V_t \) and phase angle \( \theta_t \), used to generate three-phase voltage waveform \( v^{abc}_t(j) \) for PLL frequency estimation \cite{ref23}:

\begin{equation}
f_t(j) = \frac{1}{2\pi} \frac{d\theta_t(j)}{dt},
\end{equation}

When time steps differ between transmission (\( T_T = 10 \)~ms) and distribution (\( T_D = \SI{100}{\micro\second} \)), voltage magnitude \( V_t(j) \) and phase angle \( \theta_t(j) \) remain constant throughout the 10~ms interval. However, during this period, the distribution system generates three-phase voltage waveforms \( v^{abc}_t(j) \) at every 100~$\mu$s interval using these constant magnitude and angle values. While the voltage magnitude and angle data remain static, the instantaneous three-phase voltage waveforms continue to evolve at the fine timestep resolution. This mismatch between the coarse transmission data update rate and the fine distribution waveform generation leads to distorted frequency estimation by the PLL, thereby degrading DPV frequency response performance.

\begin{figure}[!t]
\centering
\includegraphics[width=1\columnwidth]{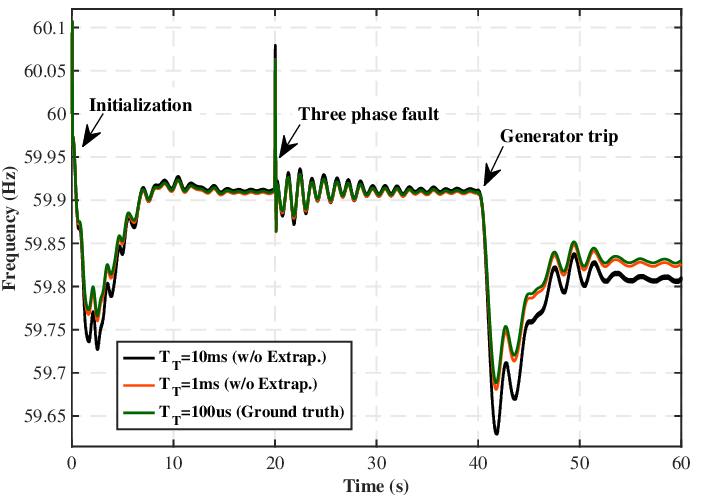}
\caption{Effect of different simulation time steps on PLL frequency calculation}
\label{fig:timesteps}
\vspace{-1em}
\end{figure}

As shown in Figure~\ref{fig:timesteps}, we compare PLL performance under three transmission timesteps: \SI{100}{\micro\second}, 1~ms, and 10~ms, with \SI{100}{\micro\second} as ground truth. At 10~ms, repeated voltage inputs over 100 cycles cause substantial PLL inaccuracies, evaluated under two disturbances: a three-phase fault at 20 seconds (0.08 seconds duration) and a 40~MW generator trip at Bus 30 at 40 seconds. The more pronounced frequency dip observed in the black line ($T_T$ = 10~ms without extrapolation) is caused by PLL error propagation. When 10~ms data is applied to the \mbox{\SI{100}{\micro\second}} distribution network without extrapolation, the voltage magnitude and phase angle remain constant for 100 consecutive timesteps, causing the PLL to erroneously calculate frequency based on these repeated values. This results in artificial frequency oscillations and amplified transient responses during disturbance events.

\subsection{PV Frequency Response}
The DPV frequency response mechanism is modeled as follows:

\begin{equation}
P^{PFR}_t(j) =
\begin{cases}
\frac{(f_0 - K^{\mathrm{DB}}_{\text{UF}}) - f_t(j)}{f_0} D, & f_t(j) < f_0 - K^{\mathrm{DB}}_{\text{UF}}, \\
\frac{(f_0 + K^{\mathrm{DB}}_{\text{OF}}) - f_t(j)}{f_0} D, & f_t(j) > f_0 + K^{\mathrm{DB}}_{\text{OF}}, \\
0, & \text{otherwise},
\end{cases} \label{eq:pfr}
\end{equation}
\begin{equation}
f^{\text{sys}}_t = \frac{1}{2\pi N_{\text{Gen}}} \sum_{i=1}^{N_{\text{Gen}}} \omega_i(t), \label{eq:fsys}
\end{equation}
\begin{equation}
\text{ACE}_t = 10 B (f^{\text{sys}}_t - f_0), \label{eq:ace}
\end{equation}
\begin{equation}
P_t^{SFR(Total)} = \sum_{k \in \mathcal{K}_j} P^{SFR}_t(k) = K_p \text{ACE}_t + K_i \int_0^t \text{ACE}_t(\tau) d\tau, \label{eq:sfr_total}
\end{equation}
\begin{equation}
P^{SFR}_t(j) = \beta_k P_t^{SFR(Total)}, \label{eq:sfr}
\end{equation}
\begin{equation}
P_t^{\text{Ref}}(j) = \min\left(P^{base}_t(j) + P^{PFR}_t(j) + P^{SFR}_t(j), P_t^{\text{Min}}(j)\right), \label{eq:ref}
\end{equation}
\begin{equation}
P^{DPV}_t(j) = \max\left(\min\left(P_t^{\text{Ref}}(j), P^{MPP}_t(j)\right), P_t^{\text{Min}}(j)\right) \label{eq:dpv}
\end{equation}

where \( f_0 = 60 \) Hz is the nominal system frequency, \( K^{\mathrm{DB}}_{\text{UF}} \) and \( K^{\mathrm{DB}}_{\text{OF}} \) are the under- and over-frequency deadbands, \( f_t(j) \) is the frequency at the PCC, \( D \) is the droop gain in MW/Hz, \( B \) is the frequency bias factor (MW/0.1 Hz), \( f^{\text{sys}}_t \) is the system frequency computed, \( N_{\text{Gen}} \) is the number of generators, \( K_p \) and \( K_i \) are the proportional and integral gains of the PI controller, \( \beta_k \) is the participation factor for the \( k \)-th resource, \( P^{base}_t(j) \) is the base power derived by subtracting the reserve from the maximum power point (MPP), \( P^{MPP}_t(j) \) is the MPP output, \( P_t^{SFR(Total)}\) is the total secondary frequency response across all resources, and \( P_t^{\text{Ref}}(j) \) is the reference power for the DPV at node \( j \) \cite{ref21}, \cite{ref22}.

Equation \eqref{eq:pfr} defines the PFR, where inverter-based DPVs respond immediately to frequency deviations at the PCC using droop control. Equations \eqref{eq:fsys}-\eqref{eq:sfr} compute system frequency and SFR components. Equations \eqref{eq:ref}-\eqref{eq:dpv} define the reference and final DPV output, constrained by maximum and minimum power limits.

The DPV models in the distribution system are simulated in the EMT domain, incorporating detailed IBR models with internal inner controls, maximum power point tracking (MPPT), maximum power point estimation (MPPE), and sophisticated EMT models. These IBRs are highly sensitive to frequency variations, underscoring the need for precise frequency calculation in T\&D co-simulation to accurately capture their dynamic response.

\vspace{-1em}
\subsection{Data Extrapolation}
\begin{figure}[!t]
\centering
\includegraphics[width=1\columnwidth]{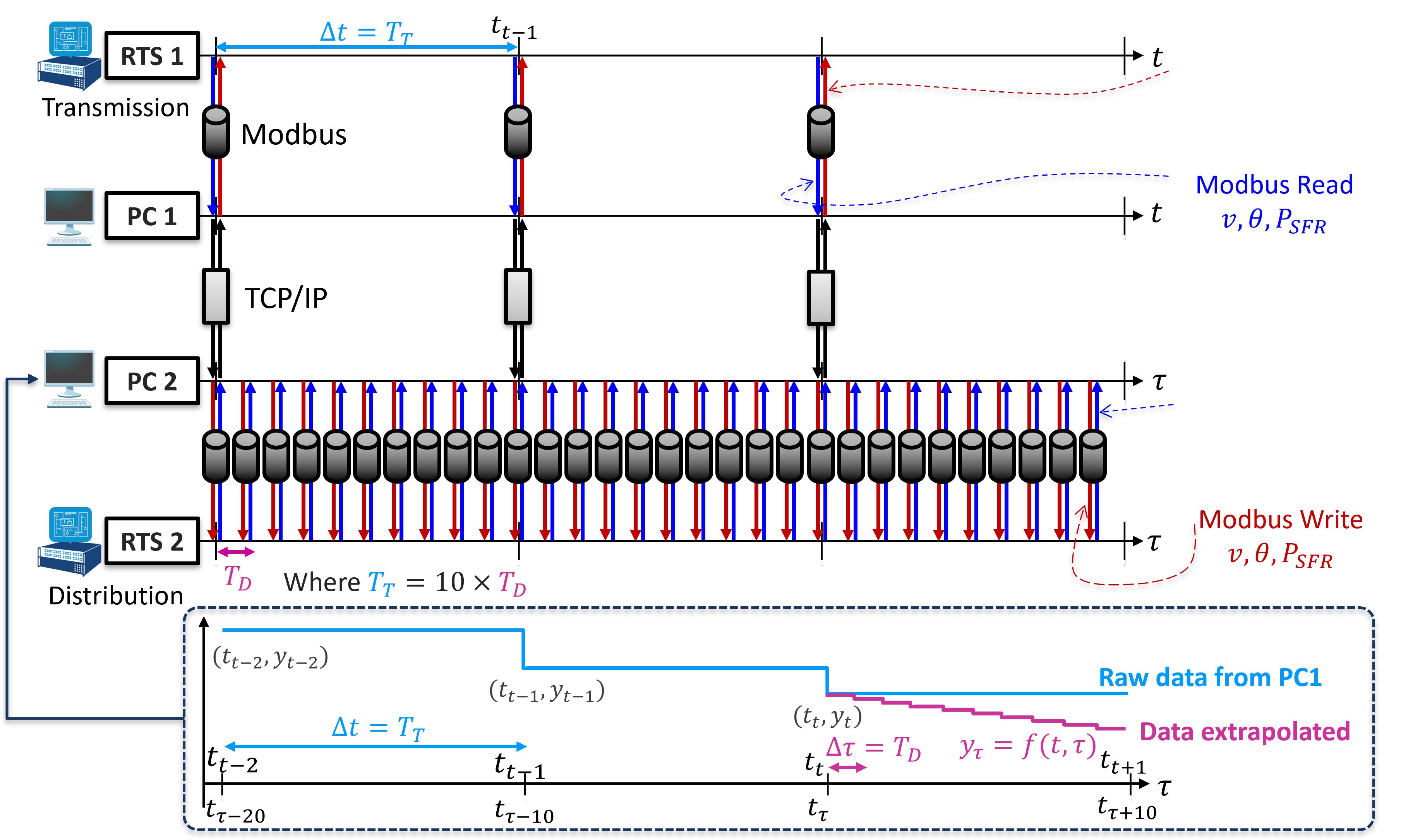}
\caption{Communication and data synchronization in T\&D co-simulation}
\label{fig:exchange}
\vspace{-1em}
\end{figure}

The T\&D co-simulation framework coordinates real-time data exchange between the IEEE 118-bus transmission system and the IEEE 123-bus distribution system, as shown in Figure~\ref{fig:exchange}, using two dedicated PCs and two real-time simulators such as OPAL-RT (one for transmission and one for distribution) with Python-based TCP/IP and Modbus protocols in a local area network. The two PCs handle data processing, logging, and extrapolation computations, providing flexibility for debugging and monitoring. TCP/IP and Modbus protocols are chosen for their wide support across different RTS platforms. While direct RTS-to-RTS communication via analog I/O or FPGA-based interfaces is possible, our middleware approach provides a more flexible and scalable solution for multi-\mbox{T\&D} scenarios.

The historical measurements refer to the three most recent voltage magnitude values ($V_{t-2}$, $V_{t-1}$, $V_t$) and phase angle values ($\theta_{t-2}$, $\theta_{t-1}$, $\theta_t$) received from the transmission system at 10~ms intervals. These values are stored in a buffer and used by the Quadratic Extrapolation method to predict intermediate values at the distribution system's 100~$\mu$s resolution. This setup transfers the magnitude of the voltage \( V_t(i) \), the phase angle \( \theta_t(i) \) and the secondary frequency response power \( P^{SFR}_t(i) \) from the transmission bus \( i \) to the distribution node \( j \), while feeding back the active power \( P_{\mathrm{t+1}} \), the reactive power \( Q_{\mathrm{t+1}} \) and the available power \( P^{avail}_{\mathrm{t+1}} \), measured at the distribution substation node, for coordinated operation \cite{ref25}. Timestep mismatches between the transmission and distribution networks, where \( T_T \) (\( \Delta T \)) is 100 times larger than \( T_D \) (\( \Delta \tau \)), lead to repeated data input in the distribution system, causing inaccuracies in PLL-based frequency calculations, as discussed in Subsection II. A. This paper addresses a 100-fold interval, but for the purpose of illustration in this figure, it is set to a 10-fold interval.

Extrapolation resolves these mismatches by predicting high-resolution data from historical measurements, filling the 100 missing timesteps between transmission updates. Three methods are evaluated: Low-Pass Filter (LPF), Linear Extrapolation, and the proposed Quadratic Extrapolation.

\subsubsection{Low-Pass Filter (LPF)}
The Low-Pass Filter (LPF) is a signal processing technique designed to smooth data by attenuating high-frequency noise, serving as a foundational method for data pre-processing \cite{ref24}. The filtering process for the distribution system is mathematically defined by the following equation:
\begin{equation}
y_\tau^{\text{LPF}} = \alpha y_\tau + (1 - \alpha) y_{\tau-1}^{\text{LPF}},
\end{equation}
where \( y_\tau \) denotes the current input from the distribution system, \( y_{\tau-1}^{\text{LPF}} \) represents the previous filtered output from the distribution system, and \( \alpha \) (a value between 0 and 1) adjusts the filter's sensitivity to new data. The parameter \( \alpha \) should be carefully tuned to balance noise reduction with the ability to track dynamic changes. For the simulation results presented in this paper, $\alpha$ values of 0.005, 0.01, 0.015, and 0.02 were tested, and $\alpha$ = 0.01 was selected as it showed the best performance.

\subsubsection{Linear Extrapolation}
Linear Extrapolation is a predictive technique that estimates future data points by projecting a linear trend based on two prior measurements, offering a computationally efficient solution for T\&D co-simulation \cite{ref25}. The method is formulated for the distribution system as follows:
\begin{equation}
y_\tau = y_t + \frac{y_t - y_{t-1}}{\Delta t} (t_\tau - t_t),
\end{equation}
where \( y_\tau \) is the predicted value at the distribution system's time (\(t_\tau\)), \( y_t \) and \( y_{t-1} \) are the current and previous values from the transmission system at time (\( t_t \)), and \( \Delta t \) is the time interval in the transmission system. The term \( t_\tau - t_t \) represents the time difference from the last transmission update to the current distribution time, but this first-order model produces non-smooth predictions.

\subsubsection{Quadratic Extrapolation (Proposed)}
Lagrange polynomials, widely used in signal processing and control systems, are first proposed here for T\&D co-simulation to address time step mismatches. The proposed Quadratic Extrapolation method uses a Lagrange polynomial with three past data points (\( y_{t-2}, y_{t-1}, y_t \)) from the transmission system at times \( t_{t-2}, t_{t-1}, t_t \):
\begin{equation}
y_\tau = \sum_{m=0}^2 y_{t-m} \prod_{\substack{n=0 \\ n \neq m}}^2 \frac{t_\tau - t_{t-n}}{t_{t-m} - t_{t-n}},
\end{equation}
where \( y_\tau \) is the predicted value for the distribution system, and Lagrange basis polynomials ensure smooth predictions. Here, \( t \) denotes the time index for the transmission system, while \( \tau \) represents the finer time index for the distribution system.

For computational efficiency, it is expressed in matrix form:
\begin{equation}
y_\tau = \mathbf{w}^T \mathbf{y}, \quad \mathbf{w} = \left( \prod_{\substack{n=0 \\ n \neq m}}^2 \frac{t_\tau - t_{t-n}}{t_{t-m} - t_{t-n}} \right)_{m=0,1,2}, 
\end{equation}
\begin{equation}
\quad \mathbf{y} = [y_t, y_{t-1}, y_{t-2}]^T,
\end{equation}
where \( \mathbf{w} \) is the weight vector, and \( \mathbf{y} \) is the vector of past data points from the transmission system, with \( \tau \) indicating the prediction point in the distribution system. 

To illustrate, consider three consecutive voltage angle data points from the transmission system: $\theta_{t-2} = -18.5^{\circ}$ at $t_{t-2} = 0$~ms, $\theta_{t-1} = -18.3^{\circ}$ at $t_{t-1} = 10$~ms, and $\theta_t = -18.0^{\circ}$ at $t_t = 20$~ms. To predict the angle at $t_\tau = 25$~ms using (11), the weights are $w_0 = 0.375$, $w_1 = -1.25$, $w_2 = 1.875$, yielding: $\hat{\theta}_\tau = 0.375(-18.5^{\circ}) + (-1.25)(-18.3^{\circ}) + 1.875(-18.0^{\circ})$ $= -17.81^{\circ}$. This extrapolated value smoothly follows the increasing trend, compared to holding $\theta_t = -18.0^{\circ}$ constant without extrapolation.

Unlike LPF, it avoids delays, and unlike Linear Extrapolation, it provides smoothness, making it ideal for modeling stable dynamics from inverter's PFR. The use of three data points allows the method to capture second-order trends, improving prediction accuracy compared to linear methods. The computational efficiency of the matrix form makes Quadratic Extrapolation suitable for real-time applications, minimizing processing overhead while maintaining high accuracy.

\vspace{-1em}
\subsection{Real-Time EWMA Threshold Adaptation (EWMA-RTTA)}
Real-time T\&D co-simulation requires robust data handling to ensure predictive accuracy in dynamic power systems, particularly when modeling the data vector \( y_t = (V_t, \theta_t) \), which represents the voltage magnitude \( V_t \) and voltage angle \( \theta_t \) at time \( t \), with data changes defined as \( \Delta y_t = y_t - y_{t-1} \). When a fault occurs in the power system, these data become discontinuous, leading to incorrect predictions based on historic trends, as existing methods fail to adapt to such abrupt changes. This discontinuity can cause even advanced techniques like the Quadratic Extrapolation method to erroneously predict continuous trends in \( V_t \) or \( \theta_t \).

To mitigate this issue, the proposed EWMA-RTTA approach monitors the magnitude of sudden data changes; if this magnitude exceeds the threshold \((TH)\), the buffer is reset, and rate limiting is applied to stabilize the data stream. The EWMA-RTTA process, integrated with the T\&D co-simulation framework, is detailed in Figure \ref{fig:flowchart_process}. When new transmission data arrives, the algorithm calculates $\Delta y_t$ and updates $TH_t$ using EWMA (Eq. 18-20). If an anomaly is detected ($|\Delta y_t| > TH_t$), the buffer is reset to prevent erroneous predictions based on pre-anomaly trends, and the system uses original data without extrapolation. If no anomaly is detected, Quadratic Extrapolation predicts voltage magnitude and phase angle at each {\SI{100}{\micro\second}} timestep. By resetting the buffer during anomalies, EWMA-RTTA prevents the propagation of errors from discontinuous data, ensuring that the predictive model remains aligned with the real-time dynamics of the T\&D system.

When a fault occurs within a 10~ms transmission interval, the transmission system captures it at the next data exchange point. The EWMA-RTTA mechanism detects this abrupt change ($\Delta y_t > TH_t$) by monitoring the magnitude of data variation and immediately resets the extrapolation buffer. This prevents error propagation from pre-fault trends. Following a detected anomaly, the distribution side uses the post-fault data directly without extrapolation until sufficient new data points (three consecutive samples) are collected to resume stable extrapolation. It should be noted that while the extrapolation predicts smooth evolution of boundary conditions, fast transients occurring within the transmission network between data exchanges are inherently limited by the co-simulation architecture---the proposed method improves frequency estimation accuracy at the distribution level but does not claim to capture sub-10ms transmission-side dynamics.

\subsubsection{Time-Invariant Threshold}
A single threshold can be determined using data collected over the entire observation period. Assuming the data follows a normal distribution, the threshold is defined as \( TH = \mu + 3\sigma \), where \( \mu \) is the mean and \( \sigma^2 \) is the variance of the distribution.
\begin{equation}
\mu = \frac{1}{N} \sum_{t=1}^N \Delta y_t,
\end{equation}
\begin{equation}
\sigma^2 = \frac{1}{N} \sum_{t=1}^N (\Delta y_t - \mu)^2,
\end{equation}
where \(\Delta y_t\) is the data change at time \(t\), \(\mu\) is the mean, \(\sigma^2\) is the variance, and \(N\) is the total number of samples (entire dataset).
If \(\Delta y_t > TH\), the data point is flagged as an outlier, triggering a buffer reset to prevent error propagation. This static method assumes a constant data distribution, limiting its ability to adapt to temporal variations in data changes, which may lead to misidentification of outliers.

\begin{figure}[!t]
\centering
\includegraphics[width=1\columnwidth]{Fig4.png}
\caption{Flowchart of the T\&D co-simulation process with EWMA-RTTA and Quadratic Extrapolation, including \SI{100}{\micro\second} internal simulation and 10 ms data exchange between PC1 and PC2.}
\label{fig:flowchart_process}
\end{figure}

\subsubsection{Time-Varying Threshold Calculated by Moving Window}
Time-Varying Thresholds, \(TH_t = \mu_t + 3\sigma_t\), can be calculated over a 1-second sliding window. If a window contains \(\omega \) data points, \(\Delta y_{t-\omega+1}, \ldots, \Delta y_t\),
the mean \(\mu_t\) and variance \(\sigma_t^2\) can be calculated as:
\begin{equation}
\mu_t = \frac{1}{\omega} \sum_{n=t-\omega+1}^t \Delta y_n,
\end{equation}
\begin{equation}
\sigma_t^2 = \frac{1}{\omega} \sum_{n=t-\omega+1}^t (\Delta y_n - \mu_t)^2,
\end{equation}
where \(\Delta y_n\) is the data change at time \(n\).

If \(\Delta y_t > TH_t\), the buffer is reset. This method adapts to recent data but is constrained by the fixed window size, which may include outdated samples, reducing responsiveness to abrupt data shifts.

\subsubsection{Time-Varying Threshold Calculated by Exponentially Weighted Moving Average}
The EWMA-RTTA method employs an exponentially weighted moving average (EWMA) to establish adaptive thresholds for detecting data changes, offering significant advantages over traditional moving average (MA) approaches. Unlike MA, which relies on a fixed window size (e.g., 1 second, corresponding to 100 samples at a 10 ms sampling rate) that requires redefinition for different resolution mismatches (e.g., 50-fold, 100-fold, or 200-fold), EWMA operates recursively without a fixed window, automatically adapting to any data resolution mismatch. By dynamically adjusting the smoothing factor \(\alpha_t\), EWMA emphasizes recent data changes, enabling it to more effectively capture real-time data trends and adapt to varying data variability.

The smoothing factor \(\alpha_t\) is dynamically adjusted based on data variability as:
\begin{equation}
\alpha_t = \min\left(0.01, \frac{|\Delta y_t - \mu_{t-1}|}{c \sigma_{t-1} + \epsilon}\right),
\end{equation}
where \(\Delta y_t = y_t - y_{t-1}\) represents the data change from time \(t-1\) to \(t\), \(\mu_{t-1}\) is the previous mean, \(\sigma_{t-1}\) is the previous standard deviation, \(c = 3\) is a scaling constant, and \(\epsilon = 10^{-6}\) prevents division by zero. The upper bound of 0.01 was empirically determined to balance responsiveness and stability, preventing overreaction to outliers while ensuring timely adaptation to legitimate data trends. The mean \(\mu_t\) and variance \(\sigma_t^2\) are updated recursively as:
\begin{equation}
\mu_t = \alpha_t \Delta y_t + (1-\alpha_t) \mu_{t-1},
\end{equation}
\begin{equation}
\sigma_t^2 = \alpha_t (\Delta y_t - \mu_t)^2 + (1-\alpha_t) \sigma_{t-1}^2,
\end{equation}
with initial conditions \(\mu_0 = 0\) and \(\sigma_0^2 = 0\). These recursive updates allow the threshold to adapt to changing data patterns. The threshold is set as:
\begin{equation}
TH_t = \mu_t + 3\sigma_t,
\end{equation}
where \(\mu_t\) is the EWMA mean, \(\sigma_t^2\) is the variance, \(\sigma_t\) is the standard deviation, and \(TH_t\) is the time-varying threshold following the 3-sigma rule. If \(\Delta y_t > TH_t\) or \(\Delta y_t < -TH_t\), the buffer is reset to mitigate error propagation.

This EWMA-based approach outperforms static and fixed-window MA methods by providing greater flexibility and responsiveness to data trends. While MA requires a predefined window size that may fail to capture rapid changes in high-IBR scenarios, EWMA's exponential weighting ensures that recent data influences the threshold more significantly, enhancing adaptability. The stability metric \((S_{\text{EWMA-RTTA}})\) is defined as:
\begin{equation}
S_{\text{EWMA-RTTA}} = 1 - \frac{\sum_{t=1}^N H(|\Delta y_t| > TH_t)}{\sum_{t=1}^N H(\Delta y_t)},
\end{equation}
where \(H\) is the indicator function, evaluating the proportion of non-outlier data points to quantify the method's effectiveness in maintaining data stability. Values close to 1.0 indicate stable operation. Furthermore, the Quadratic Extrapolation error is theoretically bounded by $O(\Delta t^3)$ based on Lagrange polynomial interpolation theory, ensuring that the extrapolation error decreases rapidly as the timestep decreases.

\section{Simulation Results}
The simulation environment, implemented at NCSU, employs two Real-Time Simulators (One RTS for the IEEE 118-Bus transmission system and another RTS for the IEEE 123-Bus distribution system) with a 10 ms transmission time step and a {\SI{100}{\micro\second}} distribution time step. The PLL implementation uses Simulink's standard SRF-PLL block with gains determined via automatic tuning at initialization: proportional gain $K_p \approx 180$, integral gain $K_i \approx 3200$, corresponding to an approximate bandwidth of 30~Hz. These gains remain fixed throughout the simulation. It should be noted that since the proposed extrapolation operates on voltage magnitude and phase angle data before they enter the PLL, the extrapolation accuracy is independent of specific PLL bandwidth settings.

The IEEE 123-Bus distribution network integrates distributed DPVs with a total capacity of 1.25 MW, providing a 200 kW reserve for frequency regulation. Data exchange between simulators is facilitated by Python-based TCP/IP and Modbus protocols in a local LAN for low-latency communication. The simulation, conducted over a 60-second period, evaluates system response under two transient events: a three-phase short-circuit fault at 20 seconds (duration 0.08 seconds) and a 40 MW generator trip at Bus 30 at 40 seconds.

Performance is assessed using MAPE and nMAE, defined as:
\begin{equation}
\text{MAPE} = \frac{1}{N} \sum_{i=1}^N \left| \frac{y_i - \hat{y}_i}{y_i} \right| \times 100,
\end{equation}
\begin{equation}
\text{nMAE} = \frac{1}{N} \sum_{i=1}^N \frac{|y_i - \hat{y}_i|}{\max(|y_i|)},
\end{equation}
where \( y_i \) is the actual value, \( \hat{y}_i \) is the predicted value, and \( N \) is the number of samples.

\begin{figure*}[!t]
\centering
\includegraphics[width=2\columnwidth]{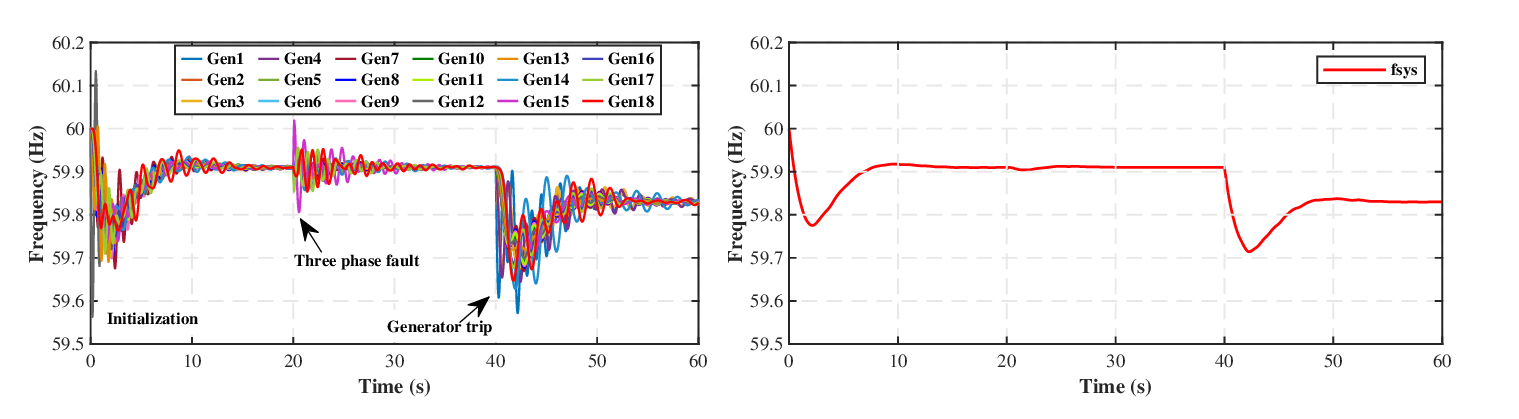}
\caption{Left: Frequencies of 18 generators in the IEEE 118-Bus system; Right: System frequency as the average of these generators.}
\label{fig:118bus_gen_freq}
\end{figure*}

Figure \ref{fig:118bus_gen_freq} illustrates the frequency dynamics of the IEEE 118-Bus system, calculated from the angular velocities of each generator. Figure \ref{fig:118bus_gen_freq} also displays the frequencies derived from the angular velocities of the 18 generators in the IEEE 118-Bus system, showing significant oscillations that vary between each generator bus due to differences in load conditions, network topology, and individual generator characteristics. The right side presents the system frequency, calculated as the average of these generator frequencies, resulting in a smoother profile that mitigates the pronounced oscillations.

\subsection{Evaluation of Voltage Magnitude and Angle Prediction}
The proposed Quadratic Extrapolation method is compared with LPF and Linear Extrapolation at a 10 ms transmission time step against the \SI{100}{\micro\second} ground truth, with the baseline method defined as a simulation where both transmission and distribution networks operate at a uniform \SI{100}{\micro\second} time step, exchanging information seamlessly. The evaluation focuses on voltage magnitude and angle predictions, which are critical for accurate PLL-based frequency calculation in T\&D co-simulation. The proposed method leverages high-resolution data to ensure smooth predictions, mitigating errors caused by time step mismatches.

\begin{figure}[!t]
\centering
\includegraphics[width=1\columnwidth]{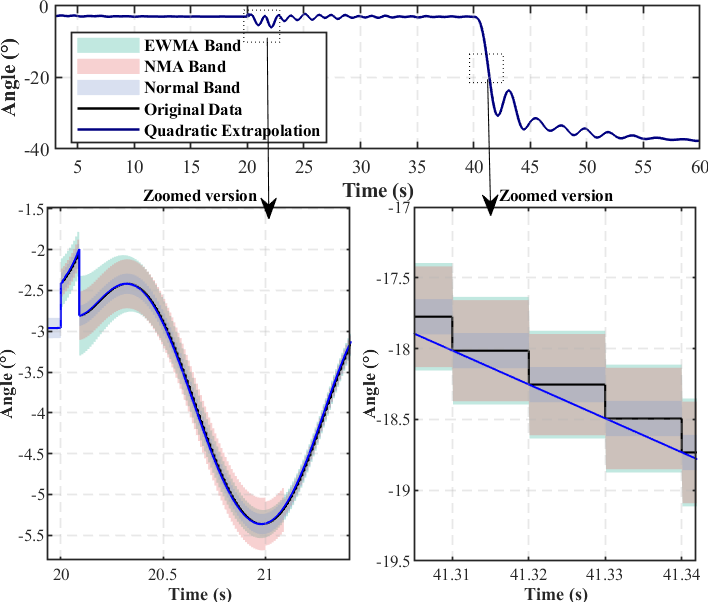}
\caption{Comparison of threshold bands for detecting anomalies in voltage angle}
\label{fig:band_comparison}
\vspace{-1em}
\end{figure}

\begin{figure}[!t]
\centering
\includegraphics[width=1\columnwidth]{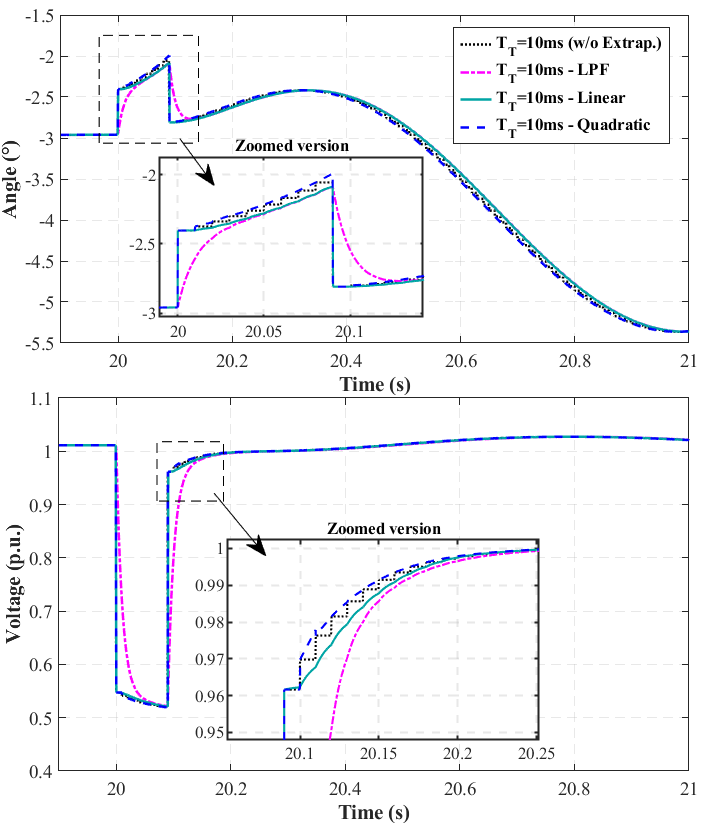}
\caption{LPF vs. Linear vs. Quadratic comparison for voltage angle (Top) and magnitude (Bottom)}
\label{fig:perf_comparison}
\vspace{-1em}
\end{figure}

Figure \ref{fig:band_comparison} evaluates threshold bands for anomaly detection in voltage and angle predictions, while Figure \ref{fig:perf_comparison} visually compares the performance of extrapolation methods over time. The performance of the proposed EWMA-RTTA method is assessed and compared with other approaches, including Normal Distribution and Moving Average (MA), using the \( S_{\text{EWMA-RTTA}} \) stability metric, as detailed in Table \ref{tab:band comparison}. 

Over a 60-second simulation with three anomaly events—three-phase short-circuit fault, fault clearance, and generator trip—the analysis shows that the Normal Distribution method estimates only two outliers for voltage data and excessively predicts 137 outliers for angle data, failing to accurately detect anomalies. 

In contrast, the proposed EWMA-RTTA method aligns most closely with the ground truth of three outliers, exhibiting \( S_{\text{EWMA-RTTA}} \) values most similar to those of the ground truth, suggesting its superior ability to identify anomalies compared to MA and other methods. These findings are supported by the data in the Table \ref{tab:band comparison}. The difference in outlier counts between voltage magnitude and angle reflects the different statistical characteristics of each variable.

\begin{table}[h]
\centering
\fontsize{8}{8}\selectfont
\renewcommand{\arraystretch}{1.2}
\caption{Comparison of Outlier Counts and $S_{\text{EWMA-RTTA}}$ Values}
\begin{tabular}{>{\raggedright\arraybackslash}p{2.5cm}cccc}
\toprule
Method & \multicolumn{2}{c}{Outlier Count} & \multicolumn{2}{c}{$S_{\text{EWMA-RTTA}}$} \\
\cmidrule(lr){2-3} \cmidrule(lr){4-5}
 & V. Mag. & V. Angle & V. Mag. & V. Angle \\
\midrule
Ground Truth & 3 & 3 & 0.9995 & 0.9995 \\
Normal Dist. (II.D.1) & 2 & 137 & 0.99967 & 0.97717 \\
Moving Avg. (II.D.2) & 20 & 69 & 0.99667 & 0.9885 \\
\textbf{EWMA-RTTA (Proposed)} & \textbf{5} & \textbf{15} & \textbf{0.99917} & \textbf{0.9975} \\
\bottomrule
\end{tabular}
\label{tab:band comparison}
\end{table}

In Figure \ref{fig:perf_comparison}, for voltage magnitude, the black line (\( T_{T}=10\mathrm{ms} \)) exhibits pronounced oscillations and deviations from the ground truth, reflecting the poor performance of sparse data without extrapolation. The pink line (\( T_{T}=10\mathrm{ms}\text{-}\mathrm{LPF} \)) shows a delayed response due to smoothing, the cyan line (\( T_{T}=10\mathrm{ms}\text{-}\mathrm{Linear} \)) produces oscillatory predictions, and the blue line (\( T_{T}=10\mathrm{ms}\text{-}\mathrm{Quadratic} \)) achieves the smoothest and most accurate trajectory, closely aligning with the ground truth. 

Similarly, for voltage angle, the black line (\( T_{T}=10\mathrm{ms} \)) shows significant errors and oscillations due to sparse sampling, the pink line (\( T_{T}=10\mathrm{ms}\text{-}\mathrm{LPF} \)) exhibits a lagged response, the cyan line (\( T_{T}=10\mathrm{ms}\text{-}\mathrm{Linear} \)) shows oscillations, and the blue line (\( T_{T}=10\mathrm{ms}\text{-}\mathrm{Quadratic} \)) provides the most stable and accurate prediction. Such errors, resulting from inaccurate predictions, can lead to inaccurate frequency calculations in PLL systems, potentially causing synchronization issues. This underscores the importance of precise frequency modeling to ensure reliable PLL performance in distribution networks.

\begin{table}[h]
\centering
\fontsize{8}{8}\selectfont
\renewcommand{\arraystretch}{1.5}
\caption{MAPE (\%) Comparison for Voltage and Angle (\SI{100}{\micro\second})}
\begin{tabular}{lcc}
\toprule
Method & V. Mag. & V. Angle \\
\midrule
$T_{T}=10\text{ms}$ (LPF) & 0.0513 & 0.408 \\
$T_{T}=10\text{ms}$ (Linear) & 0.0107 & 0.371 \\
\textbf{$T_{T}=10\text{ms}$ (Proposed)} & \textbf{0.0104} & \textbf{0.307} \\
\bottomrule
\end{tabular}
\label{tab:voltage_angle}
\vspace{-1em}
\end{table}

\vspace{-1em}
\subsection{Assessment of PLL Frequency Calculation Accuracy in T\&D Co-Simulation}
The performance of PLL is significantly affected by the mismatch between transmission time steps (\( T_{T} \)) and distribution time steps (\( T_{D} \)), especially with \( T_{T}=10 \mathrm{ms} \) data transmitted to a \( T_{D}=\SI{100}{\micro\second} \) network. This discrepancy causes oscillations and errors in PLL output, degrading frequency calculation accuracy. Reducing \( T_{T} \) to \( 1 \mathrm{ms} \) lessens repetition but fails to eliminate oscillations. To address this, three extrapolation methods—LPF, Linear, and the proposed Quadratic—are evaluated, alongside a \( \SI{100}{\micro\second}\) baseline as ground truth.

\begin{figure}[!t]
\centering
\includegraphics[width=1\columnwidth]{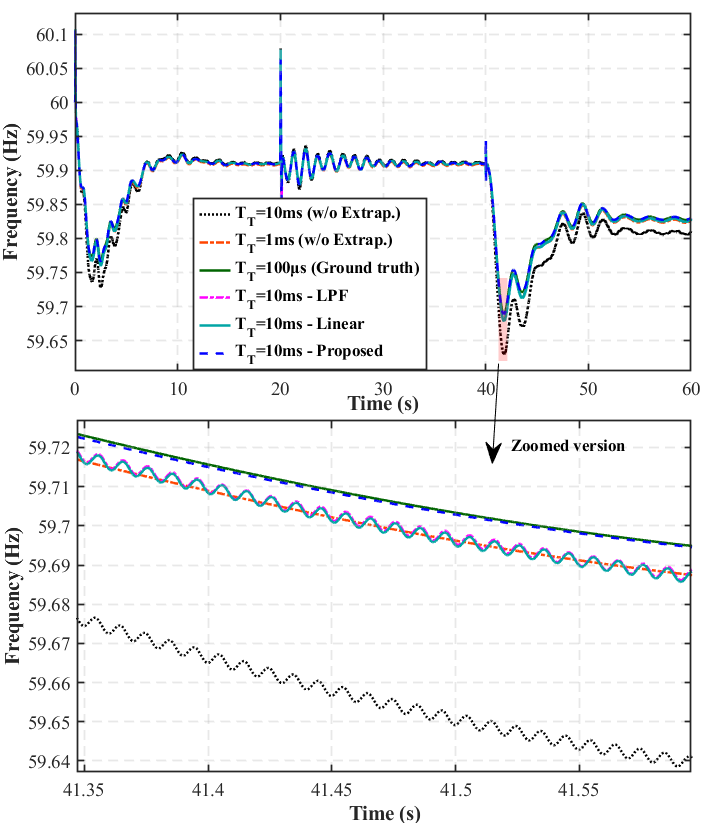}
\caption{PLL performance comparison}
\label{fig:pll_perf}
\vspace{-1em}
\end{figure}

Figure \ref{fig:pll_perf} visually compares PLL performance across these scenarios over a 60-second period, with a zoomed-in view around 41.3 to 42.1 seconds to highlight transient behavior. 

The dark green line (\( T_{T}=\SI{100}{\micro\second} \)) represents the ground truth, where the transmission time step matches the distribution time step (\( T_{D}=\SI{100}{\micro\second} \)), serving as the reference for comparison. The black line (\( T_{T}=10 \mathrm{ms} \)) exhibits the most pronounced oscillations and deviations from the ground truth, reflecting the poor performance of sparse data without extrapolation when \( 10 \mathrm{ms} \) data is applied to the \( \SI{100}{\micro\second} \) distribution network. The orange line (\( T_{T}=1\mathrm{ms} \)), with a 10-fold time step mismatch compared to \( T_{D}=\SI{100}{\micro\second} \), shows a slight reduction in oscillations, indicating a modest improvement, but still falls short of acceptable accuracy. 

Among the extrapolation methods, the pink line (\( T_{T}=10 \mathrm{ms}\text{-}\mathrm{LPF} \)) introduces significant delays due to excessive smoothing, making it the least effective extrapolation approach. The cyan line (\( T_{T}=10 \mathrm{ms}\text{-}\mathrm{Linear} \)) reduces some errors but introduces non-differentiable kinks and abrupt slope changes, leading to residual oscillations. In contrast, the blue line (\( T_{T}=10 \mathrm{ms}\text{-}\mathrm{Quadratic} \)) demonstrates the smoothest trajectory and closest alignment with the ground truth, showcasing the superiority of Quadratic Extrapolation in mitigating time step mismatches.

Table \ref{tab:pll_perf} quantifies these observations, showing that the proposed Quadratic Extrapolation achieves the lowest MAPE of 0.000563\% and a 97.08\% error improvement over the (\( T_{T}=10\mathrm{ms} \)) baseline (MAPE 0.0193\%), compared to (\( T_{T}=1\mathrm{ms} \)) (MAPE 0.00450\%, 76.67\% improvement), LPF (MAPE 0.00291\%, 84.91\% improvement), and Linear Extrapolation (MAPE 0.00300\%, 84.45\% improvement).

These results highlight Quadratic Extrapolation's effectiveness in aligning with the \SI{100}{\micro\second} ground truth, surpassing LPF's smoothing-induced delays and Linear Extrapolation's oscillatory limitations, thus ensuring robust frequency calculation critical for DPV's frequency response. 

Moreover, while Table \ref{tab:voltage_angle} in Subsection III.A indicates nearly identical results between the linear and proposed method for $v$ (derived from voltage magnitude and angle), the proposed Quadratic Extrapolation significantly enhances PLL performance, as evidenced in Table \ref{tab:pll_perf}. This difference in results arises because proposed method produces a smooth profile for $v$, whereas linear extrapolation results in non-differentiable sharp points, which adversely affect PLL applications reliant on the derivative of $v$.

\begin{table}[h]
\centering
\fontsize{8}{8}\selectfont
\renewcommand{\arraystretch}{1.5} 
\caption{MAPE and Error Improvement for PLL Performance (\SI{100}{\micro\second})}
\begin{tabular}{lcc}
\toprule
Method & MAPE (\%) & Error Improvement (\%) \\
\midrule
$T_{T}=10\text{ms}$ & 0.0193 & 0 \\
$T_{T}=1\text{ms}$ & 0.00450 & 76.67 \\
$T_{T}=\SI{100}{\micro\second}$ (LPF) & 0.00291 & 84.91 \\
$T_{T}=10\text{ms}$ (Linear) & 0.00300 & 84.45 \\
\textbf{$T_{T}=10$ms (Proposed)} & \textbf{0.000563} & \textbf{97.08} \\
\bottomrule
\end{tabular}
\label{tab:pll_perf}
\vspace{-1em}
\end{table}

\subsection{Analysis of DPV Frequency Response with EWMA-RTTA-Quadratic}
This section evaluates the PV frequency response using PLL-based frequency calculations, building on the previously mentioned simulation environment and the 40~MW generator trip scenario at Bus 30, with a focus on the case where PV operates with a 200~kW headroom. AGC (Automatic Generation Control) signal requests for SFR vary by ISO (e.g., 2 seconds in PJM, 6 seconds in NYISO), but this simulation assumes real-time requests for simplicity. 

\begin{figure}[!t]
\centering
\includegraphics[width=1\columnwidth]{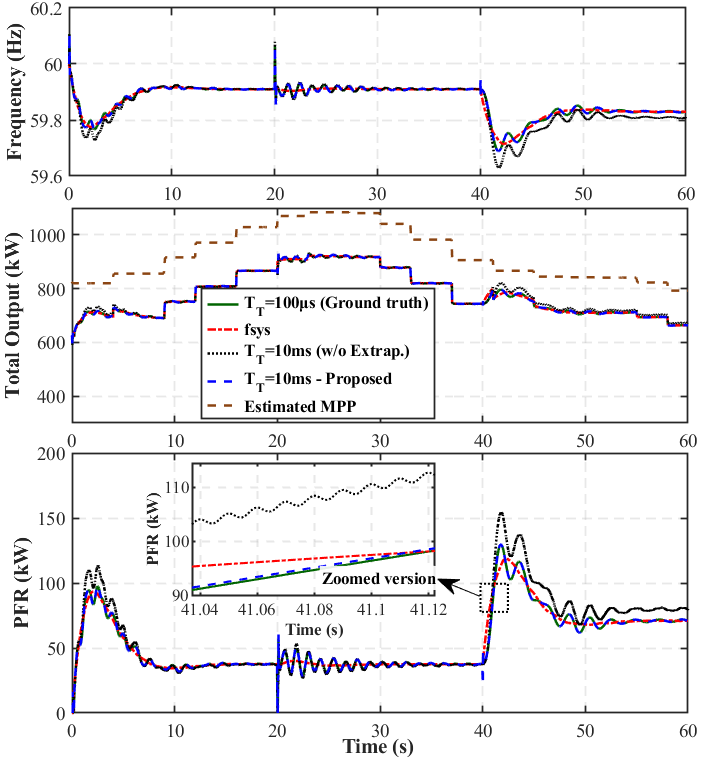}
\caption{Frequency and Distributed PV Output Comparison for Primary Frequency Response without AGC}
\label{fig:pfr_no_agc}

\vspace{1em}

\includegraphics[width=1\columnwidth]{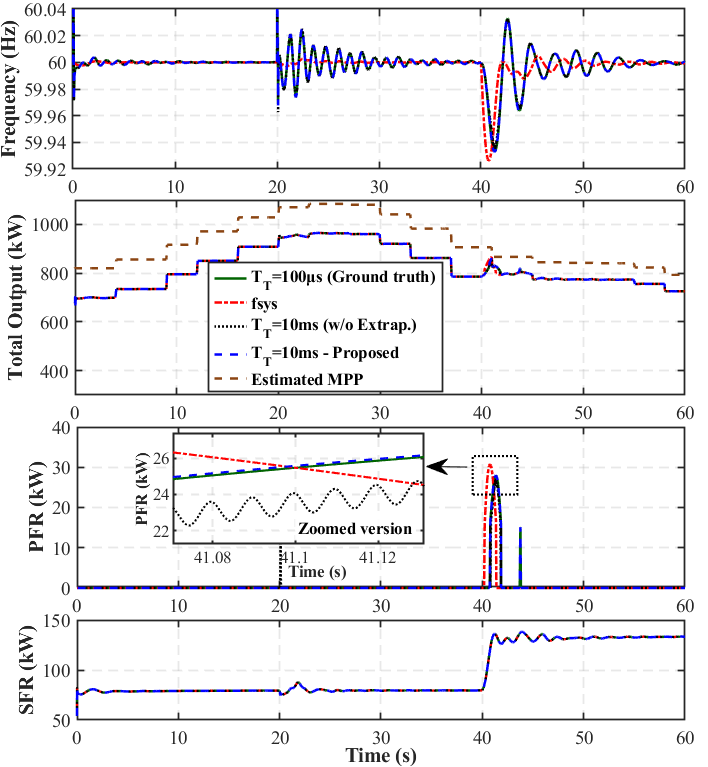}
\caption{Frequency and Distributed PV Output Comparison for Primary and Secondary Frequency Response with AGC}
\label{fig:pfr_sfr_agc}
\end{figure}

The results are presented separately for the scenario without AGC (PFR performance) and with AGC (combined PFR+SFR performance), with Figures~\ref{fig:pfr_no_agc} and~\ref{fig:pfr_sfr_agc} evaluating each respectively. The baseline method, employing uniform \(\SI{100}{\micro\second}\) time steps across transmission and distribution systems, serves as the ground truth reference for performance assessment.

\begin{figure}[!t]
\centering
\includegraphics[width=1\columnwidth]{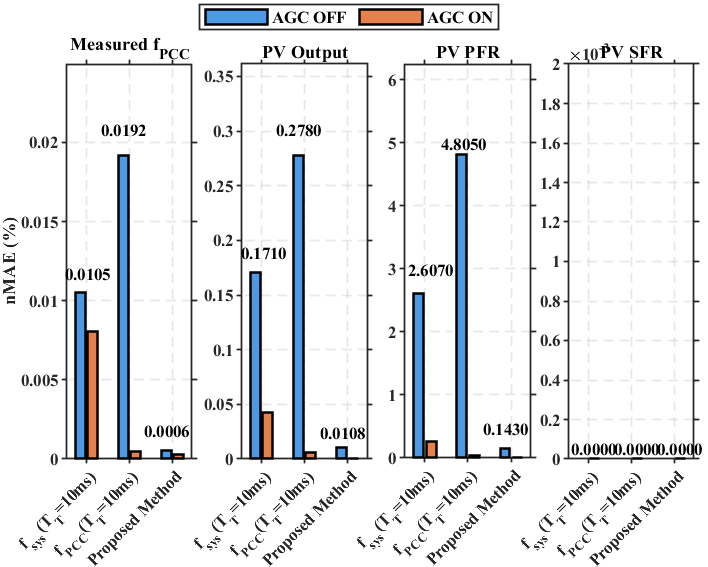}
\caption{Bar plot of nMAE (\%) comparison for different methods}
\label{fig:bars_nmae_comparison}
\end{figure}

Figures \ref{fig:pfr_no_agc} and \ref{fig:pfr_sfr_agc} assess PFR performance without AGC and combined PFR+SFR performance with AGC, respectively, for the 200 kW reserve, with quantitative nMAE results summarized in Table \ref{tab:pv_output}. In both figures, the first subplot compares frequencies, where the dark green line (\( T_{T}=\SI{100}{\micro\second} \)) represents the ground truth PLL-based calculation. The red line depicts the system frequency (\( f_{\text{sys}} \)), the black line (\( T_{T}=10\mathrm{ms} \)) reveals significant discrepancies from unextrapolated PLL application with repeated \( 10\mathrm{ms} \) data, and the blue line (\( T_{T}=10\mathrm{ms}\text{-}\mathrm{Proposed} \)) closely matches the reference. While the blue and dark green lines visually overlap due to excellent agreement, their performance differences are quantified in Figure \ref{fig:bars_nmae_comparison} and Table \ref{tab:pv_output}. 

Subsequent subplots show the corresponding DPV outputs (base and frequency response) using the same color scheme, illustrating how each frequency estimate affects PV performance; for instance, the red lines in lower plots derive from \( f_{\text{sys}} \), exposing inaccuracies relative to the ground truth, whereas blue lines demonstrate improved accuracy from the proposed method's frequency. In Figure \ref{fig:pfr_no_agc}, the PV output subplot (3,1,2) features a brown dashed line for the estimated MPP, indicating capping when \( P^{base}_t(j) + P^{PFR}_t(j) + P^{SFR}_t(j) \) exceeds MPP. Details on securing headroom for frequency response while considering the available MPP can be found in \cite{ref27}.

Table \ref{tab:pv_output} and Figure \ref{fig:bars_nmae_comparison} quantify PV plant power calculation performance for the 200 kW reserve, comparing the nMAE between approaches: substituting \( f_{\text{sys}} \) for \( f_{\text{PCC}} \) (no PLL), PLL-based \( f_{\text{PCC}} \) without mismatch mitigation (no extrapolation, \( T_{T}=10\mathrm{ms} \)), and the proposed extrapolation for accurate calculation of PLL. These quantitative results clearly demonstrate the proposed method's performance improvements, which are not visually apparent due to the close overlap with ground truth in the frequency plots. This emphasizes how frequency estimation errors propagate to PV output including PFR, underscoring the need for precise \( f_{\text{PCC}} \) over \( f_{\text{sys}} \) reliance or unaddressed mismatches.

\begin{table*}[!t]
\centering
\fontsize{8}{8}\selectfont
\renewcommand{\arraystretch}{1.5} 
\caption{Comparison of nMAE (\%) for Different Methods}
\begin{tabular}{lccccccc}
\toprule
\multirow{2}{*}{nMAE (\%)} & \multicolumn{3}{c}{AGC not applied \& FR reserve = 200} & \multicolumn{4}{c}{AGC applied \& FR reserve = 200} \\
\cmidrule(lr){2-4} \cmidrule(lr){5-8}
 & Measured $f_{\text{PCC}}$ & PV Output & PV PFR & Measured $f_{\text{PCC}}$ & PV Output & PV PFR & PV SFR \\
\midrule
$f_{\text{sys}}(T_{T}=10\text{ms})$ & 0.0105 & 0.171 & 2.607 & 0.00803 & 0.0423 & 0.264 & 0 \\
$f_{\text{PCC}}(T_{T}=10\text{ms})$ & 0.0192 & 0.278 & 4.805 & 0.000461 & 0.00578 & 0.0361 & 0 \\
\textbf{$f_{\text{PCC}}$ with proposed method} & \textbf{0.000562} & \textbf{0.0108} & \textbf{0.143} & \textbf{0.000299} & \textbf{0.000597} & \textbf{0.00373} & \textbf{0} \\
\bottomrule
\end{tabular}
\label{tab:pv_output}
\end{table*}

Without AGC, the proposed method reduces nMAE for \( f_{\text{PCC}} \) from 0.0105\% (\( f_{\text{sys}} \)) and 0. 0192\% (without extrapolation) to 0. 000562\%, for the PV output of 0. 171\% and 0. 278\% to 0. 0108\%, and for PFR from 2.607\% and 4.805\% to 0.143\%. With AGC, it decreases nMAE for \( f_{\text{PCC}} \) from 0.00803\% and 0.000461\% to 0.000299\%, for PV output from 0.0423\% and 0.00578\% to 0.000597\%, and for PFR from 0.264\% and 0.0361\% to 0.00373\%. The nMAE for SFR is maintained at 0\% in all cases, since the SFR values are directly received from the transmission system and strictly adhered to by IBR, which inherently eliminates any error. These improvements of PLL's frequency calculation, PFR, and PV output, nearing the (\SI{100}{\micro\second}) baseline, highlight the robustness of the method in future high-IBR environments for reliable grid operations. Additionally, the PFR function, equally applicable to other IBR resources such as BESS, benefits from this approach by resolving time-step mismatches. This enables an accurate calculation of the PCC frequency, significantly contributing to a proper PFR evaluation of all IBR resources throughout the system.

\subsection{Computational Overhead Analysis}
The computational overhead of each extrapolation method was measured on a desktop PC (Intel Core i9-12900K, 3.20 GHz, 64 GB RAM). Table~VI summarizes the per-timestep computational requirements. The Quadratic + EWMA method requires less than {\SI{0.5}{\micro\second}} per variable, representing approximately 0.5\% of the {\SI{100}{\micro\second}} distribution timestep ($T_D$). This overhead is negligible for real-time applications, and the proposed method provides significant accuracy improvements while maintaining computational efficiency.

\begin{table}[H]
\centering
\fontsize{8}{8}\selectfont
\renewcommand{\arraystretch}{1.2}
\setlength{\tabcolsep}{5pt}
\caption{Computational Overhead Comparison}
\begin{tabular}{l>{\centering\arraybackslash}p{2.0cm}>{\centering\arraybackslash}p{1.5cm}>{\centering\arraybackslash}p{1.5cm}}
\toprule
Method & Operations & \makecell{V. Mag.\\(\si{\micro\second})} & \makecell{V. Angle\\(\si{\micro\second})} \\
\midrule
LPF & 2 mult, 1 add & 0.263 & 0.329 \\
Linear + EWMA & \makecell{8 mult, 11 add,\\2 div, 1 sqrt} & 0.350 & 0.438 \\
\textbf{Quad. + EWMA} & \parbox{2.0cm}{\centering\textbf{13 mult, 22 add,}\\[1pt]\textbf{7 div, 1 sqrt}} & \textbf{0.427} & \textbf{0.491} \\
\bottomrule
\end{tabular}
\label{tab:computational}
\end{table}

The slightly longer computation time for voltage angle is due to additional phase wrapping operations that handle angle discontinuities at $\pm 180^{\circ}$ boundaries.

\subsection{Sensitivity Analysis for Timestep Ratios}
To demonstrate the generality of the proposed method, comprehensive sensitivity analysis was conducted across different transmission timesteps ($T_T$) and distribution timesteps ($T_D$). The ground truth is defined as the case when $T_T = T_D$ for each configuration, since the proposed extrapolation method operates within the distribution simulator and is only applicable when $T_T > T_D$. 

Tables~VII and~VIII summarize the PLL frequency MAPE results without and with AGC, respectively; the former relies solely on PFR with frequency stabilizing near 60~Hz, while the latter incorporates SFR to restore frequency to exactly 60~Hz at steady state.

\begin{table}[H]
\centering
\fontsize{7}{7}\selectfont
\renewcommand{\arraystretch}{1.1}
\caption{PLL Frequency MAPE (\%) -- Without AGC}
\begin{tabular}{ccccccc}
\toprule
$T_D \backslash T_T$ & \textbf{10ms} & 5ms & 1ms & \SI{500}{\micro\second} & \SI{100}{\micro\second} & \SI{50}{\micro\second} \\
\midrule
1ms & 0.000530 & 0.000166 & GT & N/A & N/A & N/A \\
\SI{500}{\micro\second} & 0.000551 & 0.000196 & 0.000031 & GT & N/A & N/A \\
\SI{200}{\micro\second} & 0.000563 & 0.000206 & 0.000044 & 0.000940 & N/A & N/A \\
\SI{100}{\micro\second} & 0.000563 & 0.000207 & 0.000044 & 0.000014 & GT & N/A \\
\textbf{\SI{50}{\micro\second}} & \textbf{0.000566} & 0.000209 & 0.000047 & 0.000017 & 0.000003 & GT \\
\bottomrule
\end{tabular}
\label{tab:sensitivity_no_agc}
\end{table}

\begin{table}[H]
\centering
\fontsize{7}{7}\selectfont
\renewcommand{\arraystretch}{1.1}
\caption{PLL Frequency MAPE (\%) -- With AGC}
\begin{tabular}{ccccccc}
\toprule
$T_D \backslash T_T$ & \textbf{10ms} & 5ms & 1ms & \SI{500}{\micro\second} & \SI{100}{\micro\second} & \SI{50}{\micro\second} \\
\midrule
1ms & 0.000282 & 0.000075 & GT & N/A & N/A & N/A \\
\SI{500}{\micro\second} & 0.000296 & 0.000091 & 0.000020 & GT & N/A & N/A \\
\SI{200}{\micro\second} & 0.000300 & 0.000097 & 0.000026 & 0.000007 & N/A & N/A \\
\SI{100}{\micro\second} & 0.000299 & 0.000096 & 0.000026 & 0.000007 & GT & N/A \\
\textbf{\SI{50}{\micro\second}} & \textbf{0.000300} & 0.000098 & 0.000027 & 0.000008 & 0.000002 & GT \\
\bottomrule
\end{tabular}
\label{tab:sensitivity_agc}
\end{table}

All configurations achieve MAPE below 0.001\% for PLL frequency calculation across timestep ratios ranging from 1 to 200. The result at $T_D$ = {\SI{100}{\micro\second}} and $T_T$ = 10ms (ratio = 100) shows MAPE of 0.000563\% without AGC, which is consistent with the original validation in Table~IV (0.000563\%). The consistent performance across all tested conditions demonstrates that the proposed EWMA-RTTA with Quadratic Extrapolation framework is generalizable to various {T\&D} co-simulation scenarios with different timestep mismatches without requiring parameter re-tuning.

The proposed method effectively mitigates information loss when the transmission simulation runs at a much slower timestep than the distribution simulation. As shown in Table~VIII, even with a transmission timestep of $T_T=10$~ms and a distribution timestep of $T_D=50$~\textmu s, the PLL frequency MAPE remains as low as 0.0003\%, which is well within the required simulation accuracy.

\section{Conclusion}

This study presents a framework that integrates Real-Time Threshold Adaptation using an Exponentially Weighted Moving Average (EWMA-RTTA) with quadratic extrapolation, achieving a 97\% improvement in DPV frequency response accuracy in T\&D co-simulation.

Using two real-time simulators to model the IEEE 118-bus transmission system and the IEEE 123-bus distribution system, coupled via Python-based communication protocols, the proposed method reduces the normalized mean absolute error (nMAE) in PV output modeling to 0.0108\%, assuming uniform \SI{100}{\micro\second} time steps. 

This level of accuracy is critical in high-IBR systems, where errors in frequency response modeling—particularly for local control--based primary frequency response (PFR)—can significantly distort system dynamics during contingencies, especially in future grids with PV penetration exceeding 70\%.

Primary frequency response is mandated by FERC Order~842 for all newly interconnected or modified generation resources, making accurate PFR modeling essential. The proposed framework incurs negligible computational overhead and is straightforward to implement, making it well suited for large-scale grid simulations with high IBR penetration and aligned with FERC Order~2222 requirements.

Future work will extend the framework to larger-scale T\&D co-simulation involving multiple grid interconnections and diverse distributed energy resources, including photovoltaics and energy storage systems, to further enhance grid resilience. While this paper focuses on resolving timestep mismatches, addressing communication delay and jitter in real-time co-simulation will be explored in a separate study. These efforts aim to ensure stable and reliable frequency regulation in modern power systems with high IBR penetration.

\bibliographystyle{IEEEtran}
\bibliography{references}

\end{document}